\documentclass[aps,showpacs,twocolumn,pra,superscriptaddress]{revtex4}

\usepackage[tbtags]{amsmath}
\usepackage{amssymb}
\usepackage{amsfonts}
\usepackage{graphicx}
\usepackage{caption}
\usepackage {float}
\usepackage{bmpsize}
\usepackage{mathtools}
\usepackage[colorlinks=true,citecolor=blue,urlcolor=blue]{hyperref}
\hypersetup{colorlinks,linkcolor=blue,filecolor=blue,urlcolor=blue,citecolor=blue,}
\usepackage{multirow}

\begin{document}
	
\title{
Quantum Energy Teleportation via Random Bi-Partitioning in $N-$Qubit Systems}
	
\author{Zhirong Xun}
\affiliation{Key Laboratory of Low-Dimension Quantum Structures and Quantum Control of Ministry of Education, Synergetic Innovation Center for Quantum Effects and Applications, Xiangjiang-Laboratory, Institute of Interdisciplinary Studies and Department of Physics, Hunan Normal University, Changsha 410081, China}
	
\author{Changliang Ren}\thanks{Corresponding author: renchangliang@hunnu.edu.cn}
\affiliation{Key Laboratory of Low-Dimension Quantum Structures and Quantum Control of Ministry of Education, Synergetic Innovation Center for Quantum Effects and Applications, Xiangjiang-Laboratory, Institute of Interdisciplinary Studies and Department of Physics, Hunan Normal University, Changsha 410081, China}
	
\begin{abstract}


This study investigates quantum energy teleportation (QET) using stochastic bi-partitioning in an $N-$body Hamiltonian system. In this protocol, project measurements are performed on $(N - m)$ qubits to capture quantum fluctuation information of the $N-$qubit ground state during external energy injection. Significantly, the information reaches the sites of the remaining $m$ qubits faster than the energy diffuses, allowing for extracting the ground state energy through local operations. Our results show that increasing the number of qubits $N$ enhances the available energy for QET, with efficiency peaking when $(N - 1)$ qubits are inputs and one is an output. We also find a strong correlation between energy transfer efficiency and ground-state entanglement. Increasing the parameter $\frac{k}{h}$ improves both efficiency and entanglement until reaching a plateau. Overall, more qubits lead to higher energy transfer efficiency and entanglement, highlighting their critical roles in QET performance.
\end{abstract}

\maketitle
	
\section{Introduction}
In modern physics, the vacuum is not viewed as empty space but as a dynamic energy field constantly in flux \cite{10.1093}, defined by its lowest energy level known as zero-point energy. A system at this level is referred to as being in its ground state. Generally, it is regarded as impossible to extract zero-point energy via local manipulation, as such attempts would disturb the ground state, adding energy that ultimately excites the vacuum. Nonetheless, quantum mechanics predicts the emergence of areas with negative energy density via quantum interference. This notion of negative energy has profound implications across various fields, including traversable wormholes \cite{PhysRevLett.61.1446}, quantum field theory \cite{Birrell_Davies_1982, S0217751X10049633}, and the second law of thermodynamics \cite{DAVIES1982215, ProcRSocA.364.227}. In these regions, the quantum field can possess an energy density lower than that of the vacuum's zero-point energy, provided there is sufficient positive energy elsewhere to keep the total energy above zero. Consequently, a system with zero net energy can still extract energy from vacuum fluctuations, yielding a negative energy density while maintaining overall non-negativity. Based on this idea, Hotta first proposed a quantum energy teleportation (QET) protocol for spin chain systems in 2008 \cite{JPSJ.78.034001, HOTTA20085671}. This protocol enables one party to input energy into the quantum vacuum and subsequently transmit information to another party, allowing the latter to extract that energy from the vacuum.

Since then, subsequent investigations \cite{PhysRevA.84.032336, PhysRevA.80.042323, PhysRevA.82.042329, PhysRevD.78.045006, Hotta_2010, PhysRevD.81.044025, PhysRevA.109.062208,ikeda2024} have expanded the study of QET, proposing theoretical schemes for its implementation in various quantum systems, analyzing the energy transfer capabilities of different initial states, and exploring its connection to quantum correlations. Hotta and collaborators investigated QET in quantum Hall systems \cite{PhysRevA.84.032336}, ion traps \cite{PhysRevA.80.042323}, harmonic chains \cite{PhysRevA.82.042329}, and quantum field systems \cite{PhysRevD.78.045006, Hotta_2010, PhysRevD.81.044025}. Frey \textit{et al.} analyzed QET in the Gibbs state, focusing on the role of quantum correlations \cite{Frey_2013}, while Wu \textit{et al.} proposed maximizing energy extraction using a strong local passive (SLP) state instead of the ground state \cite{PhysRevA.109.062208}. Ikeda \textit{et al.} explored the evolution of quantum correlations in the original QET protocol, using quantum discord to measure correlations in mixed states \cite{ikeda2024}. In parallel, the experimental realization and potential applications of QET have garnered growing attention \cite{PhysRevLett.130.110801, PhysRevApplied.20.024051}. In 2017, Briones proposed a theoretical solution \cite{PhysRevLett.119.050502} to increase the purity of interacting quantum systems inspired by QET protocol \cite{PhysRevA.89.012311, PhysRevA.93.022308, Trevison_2015}. In 2023, the first experimental simulation of QET was performed by using a nuclear magnetic resonance (NMR) system \cite{PhysRevLett.130.110801}, followed by independent verification on the IBM quantum cloud platform \cite{PhysRevApplied.20.024051}. In recent work, Wang \textit{et al.} have shown that both quantum energy teleportation (QET) and quantum information teleportation (QIT) \cite{PhysRevLett.70.1895} can occur simultaneously through traversable wormholes in emergent spacetime \cite{wang2024}. However, a trade-off exists between the two, as both are limited by the available entanglement resources. Xie \textit{et al.} proposed an enhanced QET protocol \cite{xie2024}, which includes an additional qubit for storing energy in a quantum register for future use, and experimentally simulated on the IBM quantum cloud platform.

However, most of the existing research is limited to two-qubit systems, using the minimal QET model designed by Hotta in 2010 \cite{HOTTA20103416}. The few studies on QET in multi-body systems are also point-to-point type \cite{PhysRevA.80.042323}, meaning the energy teleportation occurs from the $n-$th qubit to the $m-$th qubit. It is clear that QET in multi-body or complex systems is equally worthy of investigation, as ensemble-to-ensemble QET offers greater universality and application potential. In this work, we investigated quantum energy teleportation via random bi-partitioning in genuine correlated $N-$body Hamiltonian systems, which occurs from $(N - m)$ qubits to $m$ qubits. In the proposed $N-$qubit QET protocol, projective measurements are performed on any arbitrary $(N-m)$ qubits to obtain quantum fluctuation information, during which external energy is injected and treated as the system's input energy. This measurement information is then transmitted via classical communication to the remaining $m$ qubits, ensuring the transfer is faster than the diffusion speed of the injected energy. Based on this information, local operations are applied to the $m$ qubits to extract energy. Our results show that the larger the $N$, the greater the amount of energy teleported. For a fixed $N$ qubits system, the efficiency of QET increases as $m$ decreases, reaching a maximum when $(N-1)$ qubits act as inputs and one qubit acts as the output. Furthermore, the efficiency of QET is found to be correlated with the entanglement of the ground state.

	
This paper is organized as follows. In the Sec. (II), we extend the two-qubit system to the more complex $N$-body fully correlated Hamiltonian quantum model, with a detailed analysis of quantum energy teleportation within this framework and a quantitative investigation of its efficiency. In the Sec. (III), we investigate the entanglement of the ground state and analyze the relationship between the efficiency of QET and the entanglement of the ground state. The last section is for the conclusion.
\section{QET in N-body genuine correlated Hamiltonian model}

Quantum energy teleportation (QET) differs from quantum information teleportation (QIT) \cite{PhysRevLett.70.1895} by focusing on the transfer of energy rather than quantum states. In 2010, Hotta \cite{HOTTA20103416} proposed a minimal quantum energy transfer (QET) model involving two qubits. In this model, a local measurement takes place on subsystem $A$, which is far removed from subsystem $B$. Although energy will be injected into subsystem $A$ during the measurement process, if the measurement information is transmitted to subsystem $B$ through classical communication, and when the speed of information transmission exceeds the speed of energy diffusion, $B$ can extract energy from its ground state through local operations based on this information. The protocol relies on the quantum correlation between the fluctuations of $A$ and $B$, enabling the efficient extraction of zero-point energy from $B$ using operations informed by $A$. To enhance both the practical reach and theoretical depth of the QET protocol, we studied quantum energy teleportation in genuinely correlated $N-$body Hamiltonian systems via random bi-partitioning. This involves constructing QET via random bi-partitioning in $N-$body systems to explore efficient energy transfer in many-body quantum entanglement and the anticipated maximum energy extraction.

\begin{figure}[H]%
\centering
\includegraphics[width=0.48\textwidth]{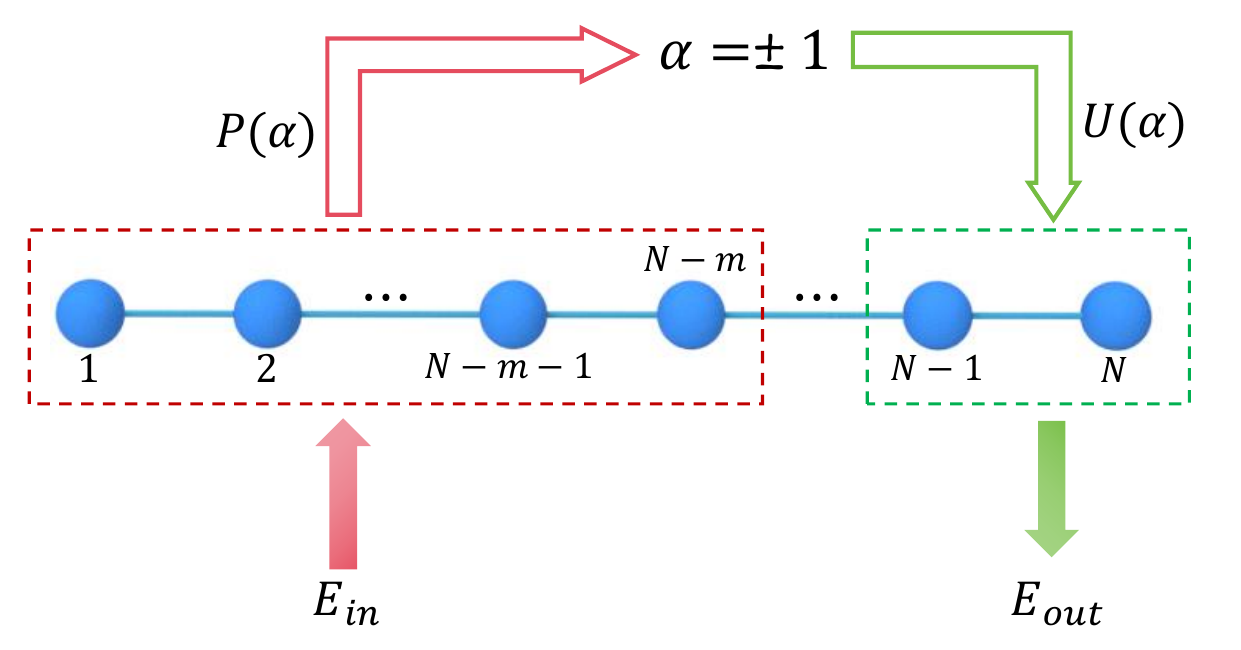}
\captionsetup{justification=justified}
\caption{\small{A schematic diagram of $(N-m)$ inputs and $m$ outputs in the QET of $N$ qubits. The qubits are denoted by circles. These qubits are numbered $1$ to $N$ from left to right. There are $(N-m)$ qubits in the red dotted box and $m$ qubits in the green dotted box. In the diagram, $P(\alpha)=(P_{1}(\alpha_{1}), P_{2}(\alpha_{2}), \dots, P_{N-m}(\alpha_{N-m}))$, $\alpha=(\alpha_{1}, \alpha_{2}, \dots, \alpha_{N-m})$, $U(\alpha)=U(\alpha_{1}, \alpha_{2}, \dots, \alpha_{N-m})$. The qubits $1$, $2$, $\dots$, $(N-m)$ are measured by the local POVM measurement defined in Eq. $(\ref{4})$, respectively, to obtain the fluctuation information $\alpha$ in the ground state $|g\rangle$. During the measurement, energy $E_{in}$ is infused into the system. The measurement results $\alpha$ are transferred to the qubits $(N-m+1), \dots, N$ through a classical channel. The energy $E_{out}$ is extracted by unitary operations $U(\alpha)$ in Eq. $(\ref{7})$ on qubits $(N-m+1), \dots, N$. }}
\label{m output}
\end{figure}

The $N$-body Hamiltonian systems that we used can be expressed as,
\begin{equation}
	\begin{aligned}
		H &=H_{1}+H_{2}+\dots+H_{N}+V \\ & =\sum_{i=1}^{N}h\sigma_{i}^{z}+2k\prod_{i=1}^{N}\sigma_{i}^{x}+\sqrt{N^{2}h^{2}+4k^{2}},
	\end{aligned}
	\label{1}	
\end{equation}
where each item is given by
\begin{align}
	H_{i}&=h\sigma _{i}^{z}+\frac{Nh^{2}}{\sqrt{N^{2}h^{2}+4k^{2}}}\quad(i\in\{1,2,\dots,N\}),\nonumber\\
       V&=2k\sigma _{1}^{x}\sigma _{2}^{x}\dots\sigma _{N}^{x}+\frac{4k^{2}}{\sqrt{N^{2}h^{2}+4k^{2}}},
       \label{2}
\end{align}
and $h$ and $k$ are positive constants with energy dimensions, $\sigma _{i}^{z}$ is the $z$-component of the Pauli operators of $i-$th qubit, the same definition for $x$. The inclusion of constant terms in Eq. $(\ref{2})$ is intended to ensure that the expected value of each operator in the ground state $|g\rangle$ is zero, which is
$\langle g| H_{i} |g\rangle=\langle g| V |g\rangle=0$. We can give the ground state of $H$ analytically (this result can be inferred through the inductive derivation from the three-qubit and four-qubit cases in the appendix),
\begin{equation}
	\begin{aligned}
		|g\rangle= & \frac{1}{\sqrt{2}}(\sqrt{1-\frac{Nh}{\sqrt{N^{2}h^{2}+4k^{2}}}}|0\rangle_{1}|0\rangle_{2}\dots|0\rangle_{N} \\ &
		-\sqrt{1+\frac{Nh}{\sqrt{N^{2}h^{2}+4k^{2}}}}|1\rangle_{1}|1\rangle_{2}\dots|1\rangle_{N}),
	\end{aligned}
	\label{3}		
\end{equation}	
where $|0\rangle$ and $|1\rangle$ is the eigenstate of $\sigma _{z}$ with eigenvalue $\pm1$.

For $N$ qubits, the QET protocol is illustrated in Fig. $\ref{m output}$, in the first step of the protocol, the projective measurements of observable $\sigma _{1}^{x}$, $\sigma _{2}^{x}$, $\dots$ and $\sigma _{N-m}^{x}$ are performed on these $(N-m)$ qubits in the ground state $|g\rangle$, respectively. The measurement results are $\alpha_{j}=\pm1$, where $j\in\{1,...,(N-m)\}$. The projective operators are represented by
\begin{equation}
{P}_{j}(\alpha_{j})=\frac{1}{2}(1+\alpha_{j}\sigma _{j}^{x})\quad(j\in\{ 1,2,\dots,(N-m)\}).
\label{4}
\end{equation}
During the measurement, the energy injected into the $j-$th qubit can be expressed as
\begin{equation}
	\begin{aligned}
		E_{j} & =\sum_{\alpha_{j}}\langle g|P_{j}(\alpha_{1})H_{j}P_{j}(\alpha_{1})|g\rangle\\
		& =\frac{Nh^{2}}{\sqrt{N^{2}h^{2}+4k^{2}}}.
	\end{aligned}	
	\label{5}		
\end{equation}
So the input energy of the subsystem can be given as
\begin{equation}
	E_{in}=(N-m)\frac{Nh^{2}}{\sqrt{N^{2}h^{2}+4k^{2}}}.
	\label{6}
\end{equation}
\begin{figure}[H]%
\centering
\includegraphics[width=0.48\textwidth]{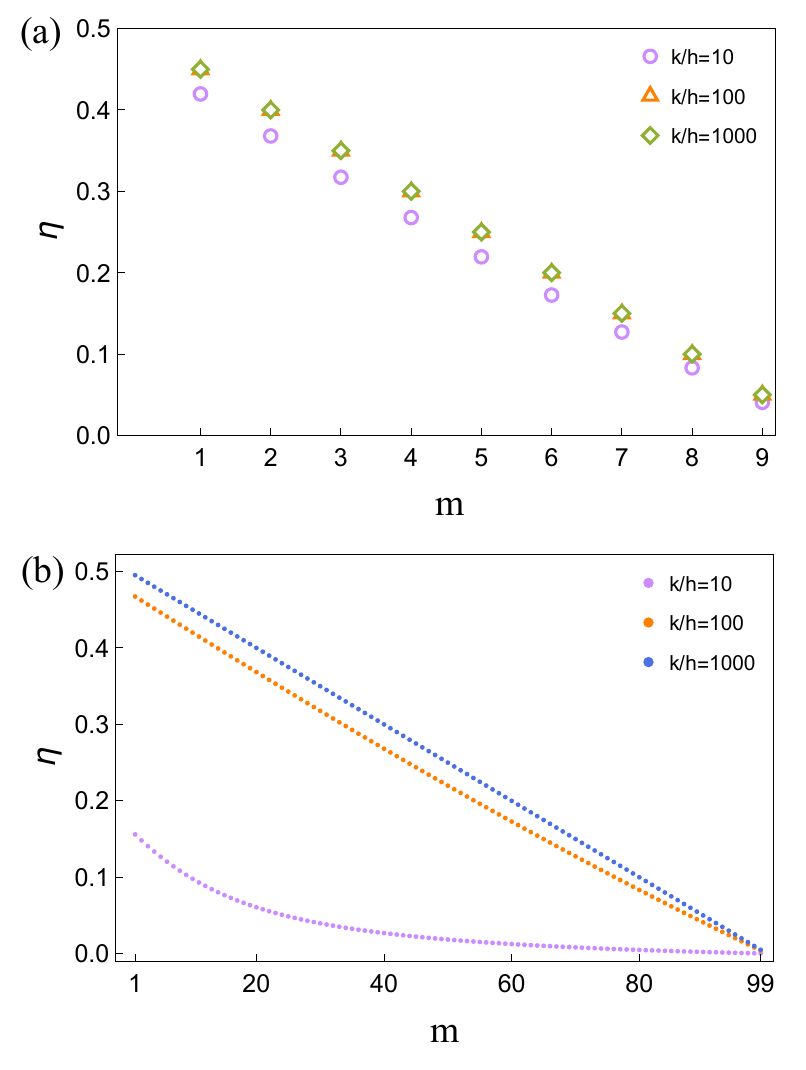}
\caption{\small{(a) When $N=10$, the energy transfer efficiency $\eta$ varies with $m$ for different $k/h$ values. (b) When $N = 100$, the energy transfer efficiency $\eta$ varies with $m$ for different $k/h$ values.}}
\label{N qubits}
\end{figure}
\noindent
Every projective operator ${P}_{j}(\alpha_{j})$
commutes with the interaction term $V$ to ensure that the energy injected during the measurement does not increase the energy of the remaining $m$ qubits. Consequently, the average energy of $H_{N-m+1}$, $\dots$, $H_{N}$ and $V$ remains zero after the measurement, matching the energy of the ground state.
		
In the second step of the protocol, the results $\alpha_{1}$, $\dots$ and $\alpha_{N-m}$
are announced to the remaining $m$ qubits through classical communication, and this process is faster than the energy diffusion speed of the system during measurement.

In the final step, a unitary operation $U(\alpha)$ is performed on the $N-m+1$, $\dots$, $N$  qubits, where its specific form depends on the value of $\alpha$. This unitary operator can be represented as
\begin{equation}
	\begin{aligned}
		U(\alpha)=\cos\theta-i\alpha\sin\theta \sigma_{N-m+1}^{y}\sigma_{N-m+2}^{x}\dots\sigma_{N}^{x},
	\end{aligned}
	\label{7}	
\end{equation}
where $\alpha=\prod\limits_{j}\alpha_{j}$.
By investigating the evolution of the density matrix, we can formulate the average state after the operation as follows
\begin{equation}
	\begin{aligned}
		\rho= &\sum_{\alpha}U(\alpha)P_{N-m}(\alpha_{N-m})\dots P_{1}(\alpha_{1}) |g\rangle\langle g| \\ & P_{1}(\alpha_{1})\dots P_{N-1}(\alpha_{N-m})U(\alpha).
	\end{aligned}
	\label{8}	
\end{equation}
On the basis of the fact that $U(\alpha)$ commutes with $H_{j}(j\in\{ 1,2,\dots,(N-m)\})$, $([U(\alpha),H_{j}]=0)$, We can compute the energy extracted from the remaining $m$ qubits as
\begin{equation}
	\begin{aligned}
		E_{out} =& \frac{1}{\sqrt{N^{2}h^{2}+4k^{2}}}[2(N-m)hk\sin2\theta \\ & -(Nmh^{2}+4k^{2})(1-\cos2\theta)].
	\end{aligned}
	\label{9}	
\end{equation}
In order to maximize $E_{out}$ with respect to $\theta$, it is easy to obtain the optimal choice for $\theta$, which satisfies
\begin{align}
\cos2\theta & =\frac{Nmh^{2}+4k^{2}}{\sqrt{(Nmh^{2}+4k^{2})^{2}+[2(N-m)hk]^{2}}},\nonumber\\
\sin2\theta & =\frac{2(N-m)hk}{\sqrt{(Nmh^{2}+4k^{2})^{2}+[2(N-
         m)hk]^{2}}}.
       \label{10}
\end{align}
Substituting $(\ref{10})$ into $(\ref{9})$ to obtain the maximum output energy $E_{out}$,
\begin{equation}
	\begin{aligned}
	E_{out}=\frac{Nmh^{2}+4k^{2}}{\sqrt{N^{2}h^{2}+4k^{2}}}[\sqrt{1+\frac{[2(N-m)hk]^{2}}{(Nmh^{2}+4k^{2})^{2}}}-1].
	\end{aligned}
    \label{11}	
\end{equation}
Hence, the energy transfer efficiency during this process can be expressed as%
\begin{equation}
	\begin{aligned}
		\eta=\frac{(Nmh^{2}+4k^{2})[\sqrt{1+\frac{[2(N-m)hk]^{2}}{(Nmh^{2}+4k^{2})^{2}}}-1]}{(N-m)Nh^{2}}.
	\end{aligned}
	\label{12}	
\end{equation}
These analytical results are the major conclusion of this work, and examples of $3$-qubit and $4$-qubit systems can be found in the appendix.

\begin{figure}[H]%
	\centering
	\includegraphics[width=0.48\textwidth]{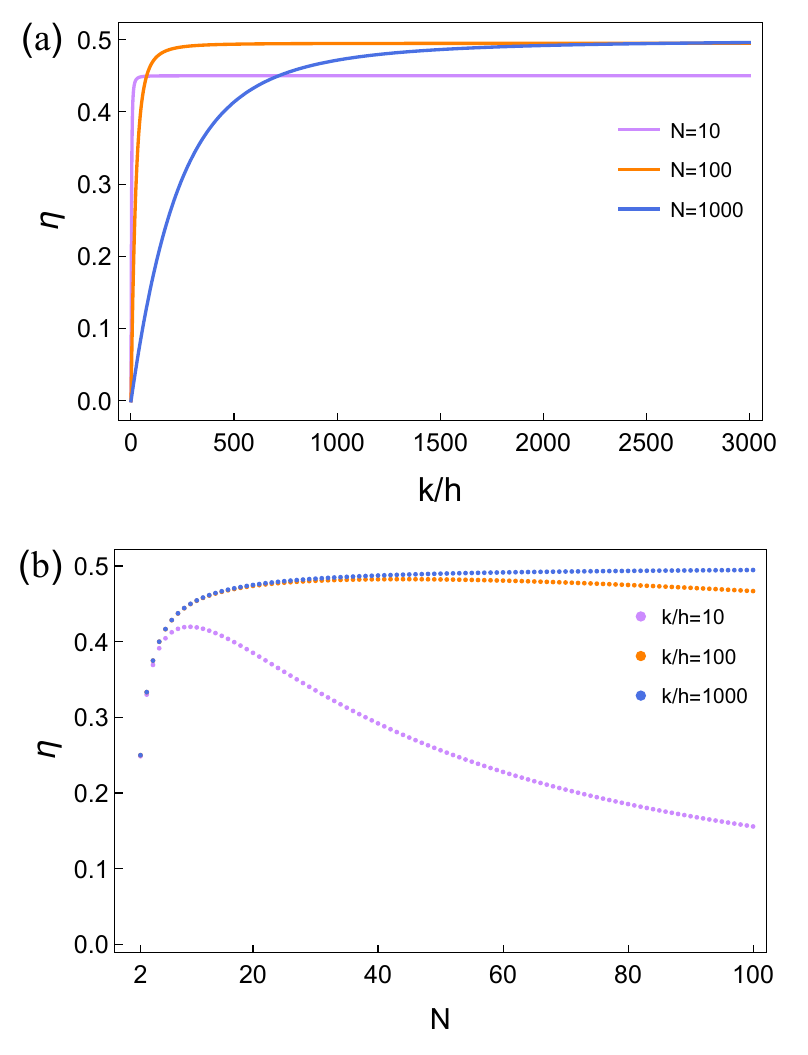}
	\caption{\small{(a) For different $N$ values, the energy transfer efficiency varies with $k/h$. (b) For different $k/h$ values, the energy transfer efficiency varies with $N$.}}
	\label{F}
\end{figure}

According to Eq. $(\ref{12})$, we illustrate the variation of energy transfer efficiency $\eta$ with respect to $m$ for different values of $k/h$ in Fig. $\ref{N qubits}$, where the number of qubits $N$ is set to $10$ and $100$, respectively. The figure indicates that, with a fixed qubit number $N$, an increase in $k/h$ enhances the energy transfer efficiency $\eta$. Notably, in Fig.~$\ref{N qubits}$(a), the overlapping curves for $k/h = 100$ and $1000$, mply that the relationship between $k/h$ and energy transfer efficiency may not follow a strictly linear trend. Comparing $(a)$ and $(b)$ in Fig.~$\ref{N qubits}$, it is evident that, with a constant parameter $N$, the energy transfer efficiency $\eta$ tends to decrease as $m$ increases. Specifically, when $k/h$ is greater than or equal to $N$, $\eta$ monotonically declines with increasing $m$, with smaller values of $m$ resulting in larger $\eta$, and $\eta$ reaching its maximum value at $m = 1$.

From the above analysis, we have established that for a given qubit number $N$, the energy transfer efficiency $\eta$ reaches its maximum when $m=1$, 
implying that the quantum energy teleportation (QET) protocol operates with only a single output in this case.
This finding motivates us to further investigate the amount of energy extracted by the QET protocol, as well as the corresponding energy transfer efficiency, under the specific condition of $m=1$.

When $m=1$ is substituted into Eqs. $(\ref{11})$ and $(\ref{12})$, the maximum teleported energy and the associated energy transfer efficiency $\eta$ for a single output are given as
\begin{align}
E_{out}&=\frac{Nh^{2}+4k^{2}}{\sqrt{N^{2}h^{2}+4k^{2}}}[\sqrt{1+\frac{[2(N-1)hk]^{2}}{(Nh^{2}+4k^{2})^{2}}}-1],
  \label{13}
 \\
\eta &\frac{(Nh^{2}+4k^{2})[\sqrt{1+\frac{[2(N-1)hk]^{2}}{(Nh^{2}+4k^{2})^{2}}}-1]}{(N-1)Nh^{2}}.
       \label{14}
\end{align}
Fig.~$\ref{F}$(a) illustrates the variation of energy transfer efficiency $\eta$ with $k/h$ for different qubit numbers $N$ (e.g., $N=10$, $N=100$, and $N=1000$) when $m=1$. The figure shows that for a given $N$, the energy transfer efficiency $\eta$ initially increases with rising $k/h$ and then stabilizes. This trend 
indicates that strengthening the interaction between qubits improves energy transfer efficiency. As $k/h$ approaches infinity, the energy transfer efficiency $\eta$ reaches different limit values depending on $N$. The results show that as $N$ increases, this limit gradually approaches $0.5$. Compared to the limit value of $0.25$ in the minimum QET model $(N=2)$, the maximum energy transfer efficiency in the N-qubit QET model is enhanced by $50\%$. Specifically, for $N=10$, the limit of $\eta$ is approximately $0.45$; for $N=100$, it increases to about $0.495$; and for $N=1000$, the limit value approaches $0.4995$. This finding emphasizes the critical role of qubit number in determining the upper bound of energy transfer efficiency.

Additionally, Fig.~$\ref{F}$(b) illustrates the variation of energy transfer efficiency $\eta$ with the number of qubits $N$ for different values of $k/h$. It is observed that for a fixed $k/h$, the energy transfer efficiency $\eta$ does not increase monotonically with $N$, but instead follows a pattern of initial increase followed by a decrease. This indicates that, in quantum systems, simply increasing the number of qubits does not always lead to improved energy transfer efficiency. For each $k/h$, there exists an optimal number of qubits, $N_{opt}$, where energy transfer efficiency is maximized. To identify this optimal $N_{opt}$, we calculate its functional relationship with $k/h$,
\begin{equation}
    \begin{aligned}
	N_{opt}=\frac{1}{2}+\frac{1}{2}\sqrt{1+C}+\frac{1}{2}\sqrt{2-C+\frac{2+16x^{2}}{\sqrt{1+C}}},
    \end{aligned}
    \label{15}	
\end{equation}
where $C=2^{4/3}(x^{2}+4x^{4})^{1/3}$ and $x=k/h$. Using this equation, we can derive $N_{opt}$ for various $k/h$ values and calculate the corresponding maximum efficiencies, which are approximately $0.42$, $0.48$, and $0.496$ for $k/h$ values of $10$, $100$, and $1000$, respectively. This highlights the importance of balancing the number of qubits $N$ with the parameter $k/h$ for optimizing energy transfer efficiency, as efficiency reaches a maximum and then rapidly decreases when $N$ significantly exceeds $k/h$.

\section{Entanglement degree of the ground state}

In the process of quantum energy teleportation, the entanglement of the ground state will be consumed. Consequently, this section investigates the relationship between the initial ground state entanglement 
with the number of qubits $N$ and the ratio $k/h$,
while analyzing how this entanglement influences energy transmission efficiency. We employ Bell's inequality violation to determine the relationship between ground state entanglement and both $N$ and $k/h$. For a subset of generalized GHZ states described by $|\psi\rangle= \cos\alpha|0,\dots,0\rangle +\sin\alpha|1,\dots,1\rangle (0\le\alpha\le\frac{\pi}{4})$, it will always violate the Bell inequality \cite{PhysRevA.74.050101}, which is
\begin{equation}
	\begin{aligned}
		\langle\mathcal{B}\rangle=(2^{N-2}\sin^{2}2\alpha+\cos^{2}2\alpha)^{\frac{1}{2}}>1 \quad(N\ge3).
	\end{aligned}
	\label{16}	
\end{equation}
That is to say, when $\langle\mathcal{B}\rangle>1$, the subset of the generalized GHZ 
states represents an entangled state, and we can judge the degree of entanglement from the degree of $\langle\mathcal{B}\rangle>1$.
\begin{figure}[H]%
	\centering
	\includegraphics[width=0.48\textwidth]{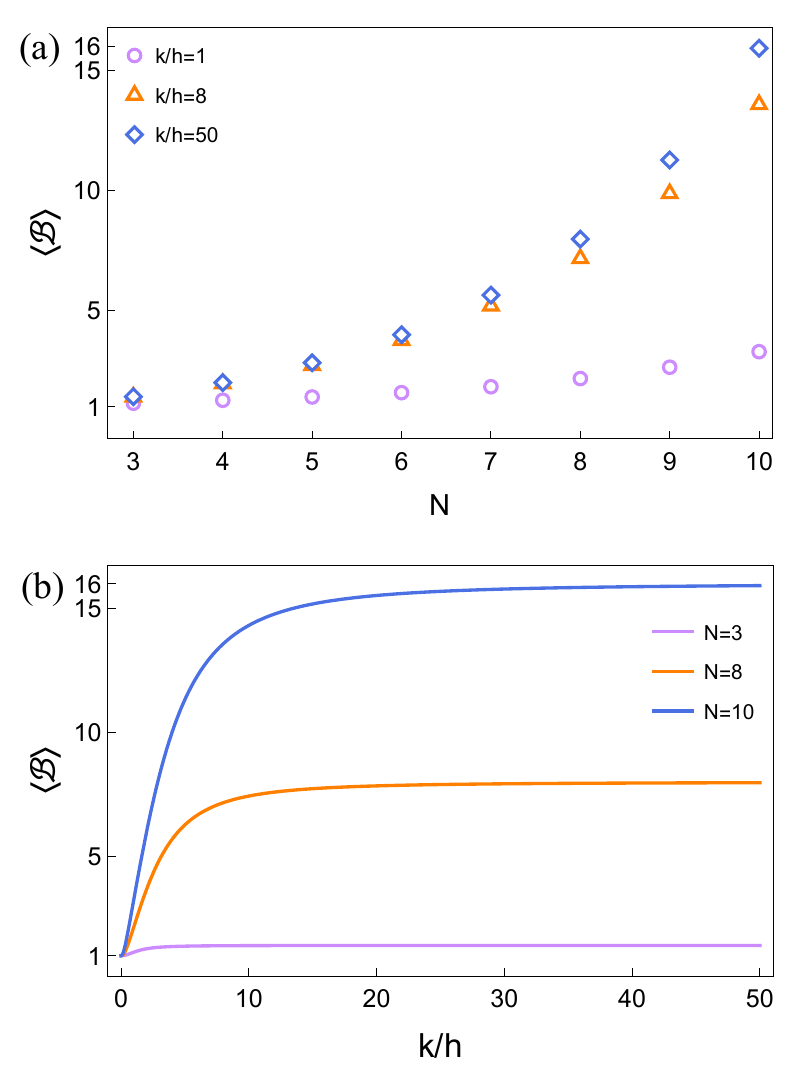}
	\caption{\small{(a) $\langle\mathcal{B}\rangle$ varies with $k/h$ for different $N$ values. (b) $\langle\mathcal{B}\rangle$ varies with $N$ for different $k/h$ values.}}
	\label{entanglement}
\end{figure}

For the ground state $|g\rangle$, $\langle\mathcal{B}\rangle$ is calculated as
\begin{equation}
	\begin{aligned}
		\langle\mathcal{B}\rangle=(2^{N-2}(\frac{-2k}{\sqrt{N^{2}h^{2}+4k^{2}}})^{2}+(\frac{-Nh}{\sqrt{N^{2}h^{2}+4k^{2}}})^{2})^{\frac{1}{2}}.
	\end{aligned}
	\label{17}	
\end{equation}
According to Eq. $(\ref{17})$, we plot the trend of $\langle\mathcal{B}\rangle$ with $k/h$ and qubit number $N$ in Fig.~$\ref{entanglement}$. It can be seen from Fig.~$\ref{entanglement}$(a) that in the case of $N\in \{3,8,10\}$ qubits, with the increase of $k/h$, $\langle\mathcal{B}\rangle$ gradually increases to the maximum and then remains unchanged, that is, there is a critical value to make $\langle\mathcal{B}\rangle$ saturated. Further analysis shows that the increase of the number of qubits $N$ increases the maximum value of $\langle\mathcal{B}\rangle$ with the increase of $k/h$. In Fig.~$\ref{entanglement}$(b), when $k/h$ is fixed, $\langle\mathcal{B}\rangle$ increases gradually with the increase of $N$, and the larger $k/h$ is, the larger the value of $\langle\mathcal{B}\rangle$ increases with $N$. Comparing the two graphs, 
it is evident that greater $N$ and $k/h$ values correspond to higher levels of ground-state entanglement.

We identified a notable correlation between the energy transfer efficiency of quantum energy transfer (QET) and the entanglement of the ground state. Specifically, when the number of qubits $N$ remains constant, increasing the system parameter $k/h$ initially enhances energy transfer efficiency and ground state entanglement, followed by a plateau phase. Further analysis reveals that as the number of qubits $N$ increases, both the maximum energy transfer efficiency and ground state entanglement of QET can be significantly improved. This finding highlights the critical role of qubit number $N$ in influencing QET performance and suggests a fundamental relationship between energy transfer efficiency and ground-state entanglement. In particular, higher ground state entanglement facilitates more efficient entanglement exchange among qubits, ultimately enhancing the system's energy transfer efficiency.

\section{Conclusion}
In this paper, we analyze the quantum energy teleportation (QET) protocol within an $N-$body fully correlated Hamiltonian model. Building on the minimum QET protocol proposed by Hotta, we extend the model from two qubits to $N$ qubits. Our investigation reveals the output energy of the $N-$body fully correlated quantum energy teleportation and analyzes the energy transfer efficiency throughout the process.

We observe that the energy transfer efficiency improves as the number of particles, $m$, for extracting energy decreases.
In particular, when there are $(N-1)$ input particles and only one output particle, the energy transfer efficiency reaches its peak. This finding suggests a direct relationship between the number of extracting particles and energy transfer efficiency in the $N-$body 
system. Notably, as the ratio $k/h$ approaches infinity and $N$ increases, the efficiency converges toward $0.5$. Moreover, the maximum energy transfer efficiency in the $N$-body system is $50\%$ higher than that in the two-body system. Additionally, we discover a strong correlation between the energy transfer efficiency of QET and the degree of ground-state entanglement, indicating that greater entanglement enhances the system's overall efficiency.

Our work explores QET through random bi-partitioning in $N-$qubit systems. Future research could involve implementing the QET protocol in Hamiltonian systems with interacting $N$ qubits, thereby broadening the protocol's application scope and enhancing experimental operability and feasibility. In the future, it is also interesting to investigate QET via random multi-partitioning in $N-$qubit systems.

\section{Acknowledgment}
		
C.R. was supported by the National Natural Science Foundation of China (Grants No. 12075245, 12421005 and No. 12247105), Hunan provincial major sci-tech program (No. 2023ZJ1010), the Natural Science Foundation of Hunan Province (2021JJ10033), the Foundation Xiangjiang Laboratory (XJ2302001) and Xiaoxiang Scholars Program of Hunan Normal University.
		

\appendix
\section{Quantum energy teleportation of three qubits}
We consider the quantum energy teleportation of three quantum systems. The system consists of three interacting qubits $A$, $B$ and $C$, the Hamiltonian of which can be expressed as,
\begin{equation}
H=H_{A}+H_{B}+H_{C}+V,
\label{A.1}
\end{equation}
where each item is given by
\begin{align}
H_{A} &=h\sigma _{A}^{z}+\frac{3h^{2}}{\sqrt{9h^{2}+4k^{2}}},
\label{A.2}	\\
H_{B} &=h\sigma _{B}^{z}+\frac{3h^{2}}{\sqrt{9h^{2}+4k^{2}}},
\label{A.3}	\\
H_{C} &=h\sigma _{C}^{z}+\frac{3h^{2}}{\sqrt{9h^{2}+4k^{2}}},
\label{A.4} \\
V &=2k\sigma _{A}^{x}\sigma _{B}^{x}\sigma _{C}^{x}+\frac{4k^{2}}{\sqrt{9h^{2}+4k^{2}}}.
\label{A.5}	
\end{align}	
The constant terms in Eqs. ($\ref{A.2}-\ref{A.5}$) are to make the expected value of each operator in the ground state $|g\rangle$ equal to zero :
$$\langle g| H_{A} |g\rangle=\langle g| H_{B} |g\rangle=\langle g| H_{C} |g\rangle=\langle g| V |g\rangle=0.$$
Since the lowest eigenvalue of the total Hamiltonian $H$ is zero, $H$ can be considered as a non-negative operator : $H\ge 0$.
The ground state of $H$ is
\begin{equation}
   \begin{aligned}
   |g\rangle=\frac{1}{\sqrt{2}}(\sqrt{1-\frac{3h}{\sqrt{9h^{2}+4k^{2}}}}|000\rangle &\\
	-\sqrt{1+\frac{3h}{\sqrt{9h^{2}+4k^{2}}}}|111\rangle).	
   \end{aligned}
   \label{A.6}	
\end{equation}
	
In three-qubit quantum energy teleportation, a projective measurement can be performed on one of the qubits and energy can be injected into it. The result of the measurement is then transferred to the other two qubits, and the positive energy is extracted from the two qubits. Or, the projective measurement can be performed on two of the qubits, energy can be injected into them, and then the measurement results are transferred to another qubit to extract positive energy from that qubit. Next, we implement the QET protocol in each of these two cases.	

\subsection{The case of one input and two outputs}
\begin{figure}[H]%
    \centering
	\includegraphics[width=0.48\textwidth]{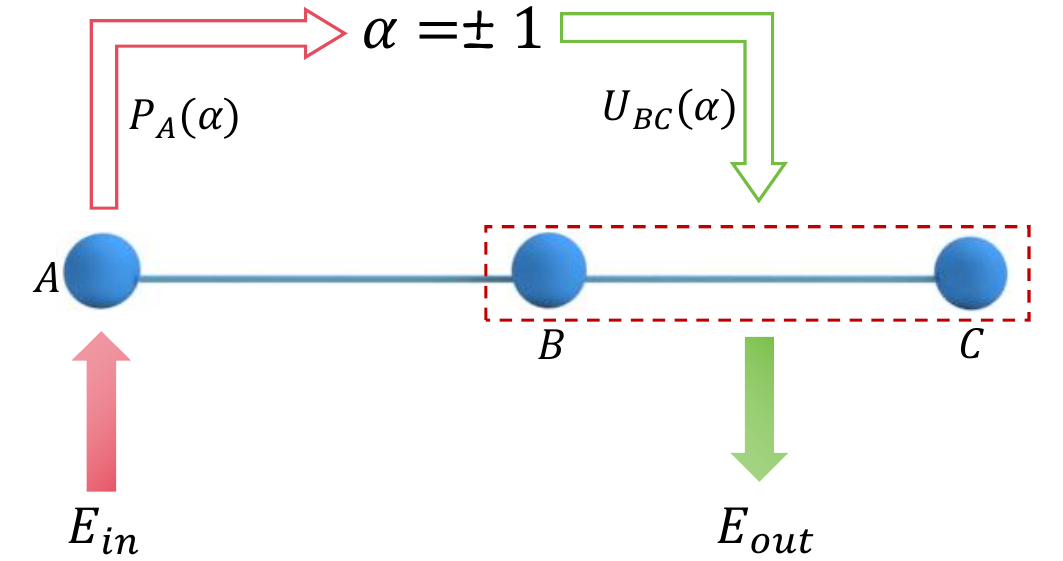}
        \captionsetup{justification=justified}
	\caption{\small{A schematic diagram of one input and two outputs in the QET of three qubits. The qubits are denoted by circles. The qubit $A$ is measured by the local POVM measurement defined in Eq. $(\ref{A.7})$ to obtain the fluctuation information $\alpha$ about $A$ in the ground state $|g\rangle$. During the measurement, energy $E_{in}$ is infused into the system. The measurement result $\alpha$ is transferred to the qubits $B$ and $C$ through a classical channel. The energy $E_{out}$ is extracted by local unitary operations $U_{BC}(\alpha)$ in Eq. $(\ref{A.9})$ on qubits $B$ and $C$.}}
	\label{two outputs}
\end{figure}

A three-qubit QET protocol with one input and two outputs is illustrated in Fig. $\ref{two outputs}$. In the first step of the protocol, a projective measurement of observable $\sigma _{A}^{x}$ is performed on $A$ in the ground state $|g\rangle$, obtaining the measurement result $\alpha=\pm1$ and the projection operator as:
\begin{equation}
{P}_{A}(\alpha)=\frac{1}{2} (1+\alpha \sigma _{A}^{x}).
\label{A.7}	
\end{equation}
During the measurement, the energy injected into $A$ is :
\begin{equation}
E_{A}=\sum_{\alpha}\langle g|P_{A}(\alpha)H_{A}P_{A}(\alpha) |g\rangle=\frac{3h^{2}}{\sqrt{9h^{2}+4k^{2}}}.
\label{A.8}	
\end{equation}
The projection operator ${P}_{A}(\alpha)$ is commutative with the interaction term $V$, $( [P_{A}(\alpha ),V]=0 )$ , to ensure that the energy injected during the measurement does not increase the energy of subsystems $B$ and $C$. Therefore, the average values of $H_{B}$, $H_{C}$ and $V$ remain zero after the measurement, which are the same as those in the ground state.
	
In the second step of the protocol, the result $\alpha$ about $A$ is announced to $B$ and $C$ through a classical channel, which is shorter than the time that the energy injected into $A$ diffuses into the system during the measurement process.
	
In the third step of the protocol, we consider a unitary operation $U_{BC}(\alpha)$ on $B$ and $C$ depending on the value of $\alpha$ given by
\begin{equation}
	U_{BC}(\alpha)=\cos\theta-i\alpha\sigma _{B}^{y}\sigma _{C}^{x}sin\theta.
	\label{A.9}	
\end{equation}
Through the evolution of the density matrix, the average state after the operation can be given by
\begin{equation}
	\rho=\sum_{\alpha}U_{BC}(\alpha)P_{A}(\alpha) |g\rangle\langle g| P_{A}(\alpha)U_{BC}(\alpha)^{\dagger}.
	\label{A.10}	
\end{equation}
On the basis of the fact that $U_{BC}(\alpha)$ commutes with $H_{A}$, $([U_{BC}(\alpha),H_{A}]=0)$, the energy extracted from $B$ and $C$ is computed as
\begin{equation}
	\begin{aligned}
		E_{out} & =E_{A}-Tr[\rho H]=-Tr[\rho(H_{B}+H_{C}+V)]\\
		& = \frac{2}{\sqrt{9h^{2}+4k^{2}}}[kh\sin2\theta-(3h^{2}+2k^{2})(1-\cos2\theta)].
	\end{aligned}
	\label{A.11}	
\end{equation}
In order to maximize $E_{out}$ with respect to $\theta$, it is easy to obtain the optimal choice for $\theta$, which satisfies
\begin{align}
\cos2\theta & =\frac{3h^{2}+2k^{2}}{\sqrt{\left(3h^{2}+2k^{2}\right)^{2}+h^{2}k^{2}}},\nonumber\\
\sin2\theta & =\frac{hk}{\sqrt{\left(3h^{2}+2k^{2}\right)^{2}+h^{2}k^{2}}}.
\label{A.12}
\end{align}
Substituting Eq. $(\ref{A.12})$ into Eq. $(\ref{A.11})$ to obtain the maximum output energy $E_{out}$,
\begin{equation}
	E_{out}= \frac{2(3h^{2}+2k^{2})}{\sqrt{9h^{2}+4k^{2}}}[\sqrt{1+\frac{h^{2}k^{2}}{(3h^{2}+2k^{2})^{2}}}-1].
	\label{A.13}	
\end{equation}
The energy transfer efficiency during this process is calculated as
\begin{equation}
	\begin{aligned}
		\eta &= \frac{E_{out}}{E_{A}}
		&=\frac{2(3h^{2}+2k^{2})(\sqrt{1+\frac{h^{2}k^{2}}{(3h^{2}+2k^{2})^{2}}}-1)}{3h^{2}} .
	\end{aligned}
	\label{1.14}
\end{equation}
\subsection{The case of two inputs and one output}
\begin{figure}[H]%
	\centering
	\includegraphics[width=0.48\textwidth]{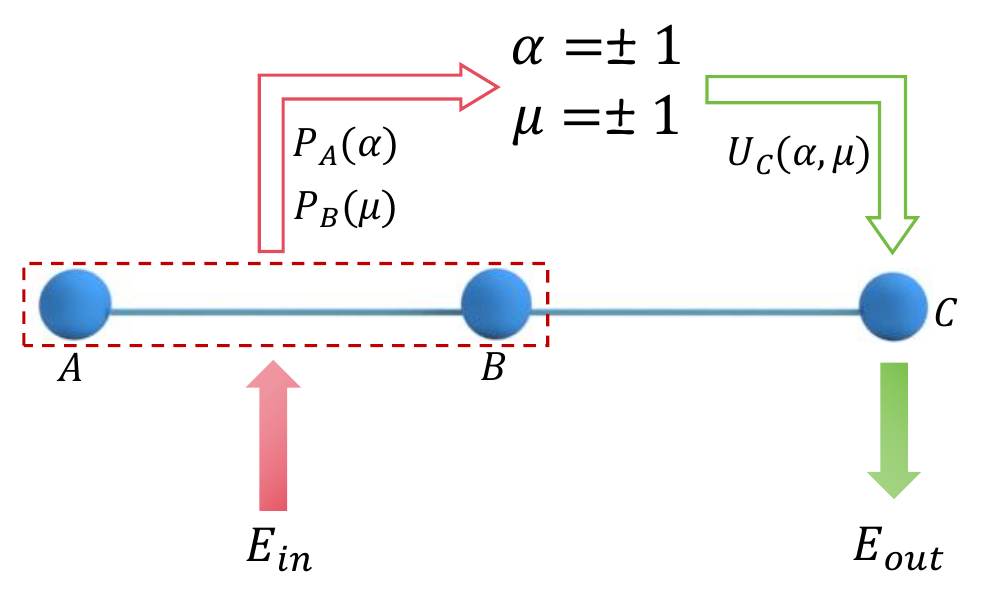}
        \captionsetup{justification=justified}
	\caption{\small{A schematic diagram of two inputs and one output in the QET of three qubits. The qubits are denoted by circles. The qubits $A$ and $B$ are measured by the local POVM measurement defined in Eqs. $(\ref{1.15})$ and $(\ref{A.16})$, respectively, to obtain the fluctuation information $\alpha$ and $\mu$ about $A$ and $B$ in the ground state $|g\rangle$. During the measurement, energy $E_{in}$ is infused into the system. The measurement results $\alpha$ and $\mu$ are transferred to the qubit $C$ through a classical channel. The energy $E_{out}$ is extracted by local unitary operations $U_{C}(\alpha, \mu)$ in Eq. $(\ref{A.20})$ on qubit $C$.}}
	\label{one output}
\end{figure}

For the case of two inputs and one output, the three-qubit QET protocol is shown in Fig. $\ref{one output}$, which is as follows. Firstly, the projective measurements of observable $\sigma _{A}^{x}$ and $\sigma _{B}^{x}$ are performed on $A$ and $B$ in the ground state $|g\rangle$, obtaining the measurement results $\alpha=\pm1$ and $\mu=\pm1$. The projection operators are given by
\begin{align}
	{P}_{A}(\alpha)&=\frac{1}{2} (1+\alpha \sigma _{A}^{x}),
	\label{1.15}\\
	{P}_{B}(\mu)&=\frac{1}{2} (1+\mu\sigma _{B}^{x}).
	\label{A.16}
\end{align}	
\noindent
During the measurement, the energy injected into $A$ is defined by
\begin{equation}
	\begin{aligned}
		E_{A} & =\sum_{\alpha,\mu}\langle g|P_{A}(\alpha)P_{B}(\mu)H_{A}P_{B}(\mu)P_{A}(\alpha)|g\rangle\\
		& =\sum_{\alpha}\langle g|P_{A}(\alpha)H_{A}P_{A}(\alpha)|g\rangle\\
		& =\frac{3h^{2}}{\sqrt{9h^{2}+4k^{2}}}.
	\end{aligned}	
	\label{A.17}	
\end{equation}
During the measurement, the energy injected into $B$ is defined by
\begin{equation}
	\begin{aligned}
		E_{B} & =\sum_{\alpha,\mu}\langle g|P_{A}(\alpha)P_{B}(\mu)H_{B}P_{B}(\mu)P_{A}(\alpha)|g\rangle\\
		& =\sum_{\mu}\langle g|P_{B}(\mu)H_{B}P_{B}(\mu)|g\rangle\\
		& =\frac{3h^{2}}{\sqrt{9h^{2}+4k^{2}}}.
	\end{aligned}	
	\label{A.18}	
\end{equation}
The energy $E_{A}$ and $E_{B}$ are regarded as the energy input from the external system via the measurement of $A$ and $B$. Therefore, the input energy of the system during the measurement process is given by
\begin{equation}
	E_{in}=\frac{6h^{2}}{\sqrt{9h^{2}+4k^{2}}}.
	\label{A.19}	
\end{equation}
	
Secondly, the results $\alpha$ and $\mu$ are announced to $C$ by a classical channel, which is shorter than the time that the energy injected into $A$ diffuses into the system during the measurement process.
	
Thirdly, we consider a unitary operation $U_{C}(\alpha, \mu)$ on C depending on the value of $\alpha$ and $\mu$ given by
\begin{equation}
	U_{C}(\alpha,\mu)=\cos\theta-i\alpha\mu\sin\theta\sigma _{C}^{y}.
	\label{A.20}	
\end{equation}
\begin{figure}[H]%
	\centering
	\includegraphics[width=0.48\textwidth]{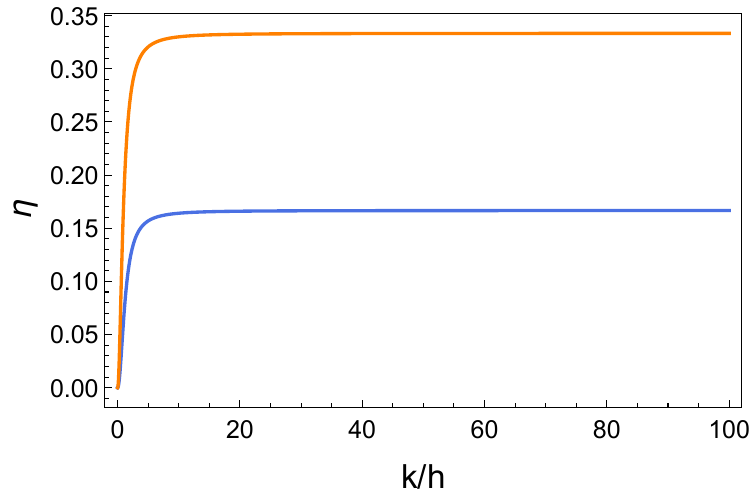}
	\caption{\small{The figure shows the change of energy transfer efficiency $\eta$ with $k/h$ for two inputs and one output (blue line) and for one input and two outputs (orange line).
		}}\label{three qubits}
\end{figure}
\noindent
Through the evolution of the density matrix, the average state after the operation can be given by
\begin{equation}
	\rho=\sum_{\alpha,\mu}U_{C}(\alpha,\mu)P_{B}(\mu)P_{A}(\alpha) |g\rangle\langle g| P_{A}(\alpha)P_{B}(\mu)U_{C}(\alpha,\mu)^{\dagger}.
	\label{A.21}	
\end{equation}
On the basis of the fact that $U_{C}(\alpha, \mu)$ commutes with $H_{A}$ and $H_{B}$, the energy extracted from $C$ is computed as
\begin{equation}
	\begin{aligned}
		E_{out} & = -\sum_{\alpha,\mu} \langle g|P_{A}(\alpha)P_{B}(\mu)U_{C}(\alpha,\mu)^{\dagger}(H_{C}+V)U_{C}(\alpha,\mu) |g\rangle \\
		& = \frac{1}{\sqrt{9h^{2}+4k^{2}}}[4kh\sin2\theta-(3h^{2}+4k^{2})(1-\cos2\theta)].	
	\end{aligned}
	\label{A.22}	
\end{equation}
In order to maximize $E_{out}$ with respect to $\theta$, it is easy to obtain the optimal choice for $\theta$, which satisfies
\begin{align}
\cos2\theta & =\frac{3h^{2}+4k^{2}}{\sqrt{(3h^{2}+4k^{2})^{2}+(4kh)^{2}}},\nonumber\\
\sin2\theta & =\frac{4hk}{\sqrt{(3h^{2}+4k^{2})^{2}+(4kh)^{2}}}.
\label{A.23}
\end{align}
Substituting Eq. $(\ref{A.23})$ into Eq. $(\ref{A.22})$ to obtain the maximum output energy $E_{out}$,
\begin{equation}
	E_{out}= \frac{3h^{2}+4k^{2}}{\sqrt{9h^{2}+4k^{2}}}[\sqrt{1+\frac{(4kh)^{2}}{(3h^{2}+4k^{2})^{2}}}-1].
	\label{A.24}	
\end{equation}
The energy transfer efficiency during this process is calculated as
\begin{equation}
	\begin{aligned}
		\eta &=\frac{E_{out}}{E_{in}} & =\frac{(3h^{2}+4k^{2})(\sqrt{1+\frac{4^{2}h^{2}k^{2}}{(3h^{2}+4k^{2})^{2}}}-1)}{6h^{2}} .
	\end{aligned}
	\label{A.25}
\end{equation}
		
Figure \ref{three qubits} shows the variation of energy transfer efficiency $\eta$ with $k/h$ in two cases. We find that the energy transfer efficiency increases with the increase of $k/h$, reaching a maximum value and then remaining near the maximum value. When $k\to\infty$, the energy transfer efficiency $\eta\to\frac{1}{6}$ in the case of one input and two outputs; the energy transfer efficiency $\eta\to\frac{1}{3}$ in the case of two inputs and one output.  Comparing the blue and orange lines, we find that the energy transfer efficiency with two inputs and one output is higher than with one input and two outputs.

\section{Quantum energy teleportation of four qubits}
We consider the quantum energy teleportation of four quantum systems.  The system consists of four interacting qubits $1$, $2$, $3$ and $4$, the Hamiltonian of which can be expressed as,
\begin{equation}
H=H_{1}+H_{2}+H_{3}+H_{4}+V,
\label{B.1}
\end{equation}
where each item is given by
\begin{align}
H_{1} &=h\sigma _{1}^{z}+\frac{2h^{2}}{\sqrt{4h^{2}+k^{2}}},\nonumber\\
H_{2} &=h\sigma _{2}^{z}+\frac{2h^{2}}{\sqrt{4h^{2}+k^{2}}},\nonumber\\
H_{3} &=h\sigma _{3}^{z}+\frac{2h^{2}}{\sqrt{4h^{2}+k^{2}}},\nonumber\\
H_{4} &=h\sigma _{4}^{z}+\frac{2h^{2}}{\sqrt{4h^{2}+k^{2}}},\nonumber\\
V &=2k\sigma _{1}^{x}\sigma _{2}^{x}\sigma _{3}^{x}\sigma _{4}^{x}+\frac{2k^{2}}{\sqrt{4h^{2}+k^{2}}},
\label{B.2}	
\end{align}	
The inclusion of constant terms in Eq. $(\ref{2})$ is intended to ensure that the expected value of each operator in the ground state $|g\rangle$ is zero, which is
$\langle g|H_{1}|g\rangle = \langle g|H_{2}|g\rangle=\langle g|H_{3} |g\rangle=\langle g|H_{4}|g\rangle=\langle g|V|g\rangle=0$. We can give the ground state of $H$ analytically,	
\begin{equation}
   \begin{aligned}
   |g\rangle=\frac{1}{\sqrt{2}}(\sqrt{1-\frac{2h}{\sqrt{4h^{2}+k^{2}}}}|0000\rangle &\\
	-\sqrt{1+\frac{2h}{\sqrt{4h^{2}+k^{2}}}}|1111\rangle).	
   \end{aligned}
   \label{B.3}	
\end{equation}	
In the four-qubit quantum energy teleportation, there are three cases in which QET protocol can be realized.
\subsection{The case of one input and three outputs}
In the first scenario, we consider one input and three outputs. Firstly, a projective measurement of observable $\sigma _{1}^{x}$ is performed on qubit $1$ in the ground state $|g\rangle$, obtaining the measurement result $\alpha=\pm1$ and the projection operator as:
\begin{equation}
\hat{P}_{1}(\alpha)=\frac{1}{2} (1+\alpha_{1}\sigma _{1}^{x}).
\label{B.4}	
\end{equation}
The total energy injected into the system during the measurement is given by:
\begin{equation}
E_{in}=\frac{2h^{2}}{\sqrt{4h^{2}+k^{2}}}.
\label{B.5}	
\end{equation}
After sending the information $\alpha$ to qubits $2$, $3$ and $4$ through classical communication, the operation $U_{234}(\alpha)$ is performed on qubits $2$, $3$ and $4$, which is defined by
\begin{equation}
U_{234}(\alpha)=\cos\theta-i\alpha\sin\theta\sigma_{2}^{y}\sigma_{3}^{x}\sigma_{4}^{x}.
\label{B.6}	
\end{equation}
The average state after the operation can be given by
\begin{equation}
\rho=\sum_{\alpha}U_{234}(\alpha)P_{1}(\alpha) |g\rangle\langle g| P_{1}(\alpha)U_{234}(\alpha)^{\dagger}.
\label{B.7}	
\end{equation}
The energy extracted from qubits $2$, $3$ and $4$ is computed as
\begin{equation}
E_{out}= \frac{1}{\sqrt{4h^{2}+k^{2}}}[hk\sin2\theta-(6h^{2}+2k^{2})(1-\cos2\theta)].	
\label{B.8}	
\end{equation}
In order to maximize $E_{out}$ with respect to $\theta$, it is easy to obtain the optimal choice for $\theta$, which satisfies
\begin{align}
\cos2\theta & =\frac{6h^{2}+2k^{2}}{\sqrt{(6h^{2}+2k^{2})^{2}+(kh)^{2}}},\nonumber\\
\sin2\theta & =\frac{hk}{\sqrt{(6h^{2}+2k^{2})^{2}+(kh)^{2}}}.
\label{B.9}
\end{align}
Substituting Eq. $(\ref{B.9})$ into Eq. $(\ref{B.8})$ to obtain the maximum output energy $E_{out}$,
\begin{equation}
E_{out}= \frac{6h^{2}+2k^{2}}{\sqrt{4h^{2}+k^{2}}}[\sqrt{1+\frac{(hk)^{2}}{(6h^{2}+2k^{2})^{2}}}-1].
\label{B.10}	
\end{equation}
The energy transfer efficiency during this process is calculated as
\begin{equation}
	\begin{aligned}
		\eta=\frac{E_{out}}{E_{in}}=\frac{(6h^{2}+2k^{2})[\sqrt{1+\frac{(hk)^{2}}{(6h^{2}+2k^{2})^{2}}}-1]}{2h^{2}}.
	\end{aligned}
	\label{B.11}
\end{equation}
\subsection{The case of two inputs and two outputs}
In the second scenario, we consider two inputs and two outputs. Firstly, the projective measurements of observable $\sigma _{1}^{x}$ and $\sigma _{2}^{x}$ are performed on qubits $1$ and $2$ in the ground state $|g\rangle$, obtaining the measurement results $\alpha_{1}=\pm1$ and $\alpha_{2}=\pm1$. The projection operators are given by
\begin{align}
\hat{P}_{1}(\alpha_{1})&=\frac{1}{2} (1+\alpha_{1}\sigma _{1}^{x}),\nonumber\\
\hat{P}_{2}(\alpha_{2})&=\frac{1}{2} (1+\alpha_{2}\sigma _{2}^{x}).
\label{B.12}
\end{align}
The total energy injected into the system during the measurement is given by:
\begin{equation}
E_{in}=\frac{4h^{2}}{\sqrt{4h^{2}+k^{2}}}.
\label{B.13}	
\end{equation}
After sending the information $\alpha_{1}$ and $\alpha_{2}$ to qubits $3$ and $4$ through classical communication, the operation $U_{34}(\alpha_{1},\alpha_{2})$ is performed on qubits $3$ and $4$, which is defined by
\begin{equation}
U_{34}(\alpha_{1},\alpha_{2})=\cos\theta-i\alpha_{1}\alpha_{2}\sin\theta\sigma_{3}^{y}\sigma_{4}^{x}.
\label{B.14}	
\end{equation}
The average state after the operation can be given by
\begin{equation}
   \begin{aligned}
    \rho=&\sum_{\alpha_{1},\alpha_{2}}U_{34}(\alpha_{1},\alpha_{2})P_{2}(\alpha_{2})P_{1}((\alpha_{1})) |g\rangle\langle g| P_{1}(\alpha_{1})\\&  P_{2}(\alpha_{2})U_{34}(\alpha_{1},\alpha_{2})^{\dagger}.	
   \end{aligned}
   \label{B.15}	
\end{equation}
The energy extracted from qubits $3$ and $4$ is computed as
\begin{equation}
E_{out}=\frac{2}{\sqrt{4h^{2}+k^{2}}}[hk\sin2\theta-(2h^{2}+k^{2})(1-\cos2\theta)].	
\label{B.16}	
\end{equation}
In order to maximize $E_{out}$ with respect to $\theta$, we can obtain the optimal choice for $\theta$, which satisfies
\begin{align}
\cos2\theta & =\frac{2h^{2}+k^{2}}{\sqrt{(2h^{2}+k^{2})^{2}+(kh)^{2}}}, \nonumber\\
\sin2\theta & =\frac{hk}{\sqrt{(2h^{2}+k^{2})^{2}+(kh)^{2}}}.
\label{B.17}	
\end{align}
Substituting Eq. $(\ref{B.17})$ into Eq. $(\ref{B.16})$ to obtain the maximum output energy $E_{out}$,
\begin{equation}
E_{out}=\frac{4h^{2}+4k^{2}}{\sqrt{4h^{2}+k^{2}}}[\sqrt{1+\frac{(kh)^{2}}{(2h^{2}+2k^{2})^{2}}}-1].
\label{B.18}	
\end{equation}
The energy transfer efficiency during this process is calculated as
\begin{equation}
	\eta=\frac{(4h^{2}+4k^{2})[\sqrt{1+\frac{(hk)^{2}}{(2h^{2}+2k^{2})^{2}}}-1]}{4h^{2}}.
	\label{B.19}
\end{equation}
\subsection{The case of three inputs and one output}
In the third scenario, we consider three inputs and one output. Firstly, the projective measurements of observable $\sigma _{1}^{x}$, $\sigma _{2}^{x}$
and $\sigma _{3}^{x}$ are performed on qubits $1$, $2$ and $3$ in the ground state $|g\rangle$, obtaining the measurement results $\alpha_{1}=\pm1$, $\alpha_{2}=\pm1$ and $\alpha_{3}=\pm1$. The projection operators are given by
\begin{align}
\hat{P}_{1}(\alpha_{1})&=\frac{1}{2} (1+\alpha_{1}\sigma _{1}^{x}),\nonumber\\
\hat{P}_{2}(\alpha_{2})&=\frac{1}{2} (1+\alpha_{2}\sigma _{2}^{x}),\nonumber\\
\hat{P}_{3}(\alpha_{3})&=\frac{1}{2} (1+\alpha_{3}\sigma _{3}^{x}).
\label{B.20}
\end{align}
The total energy injected into the system during the measurement is given by:
\begin{equation}
E_{in}=\frac{6h^{2}}{\sqrt{4h^{2}+k^{2}}}.
\label{2.21}	
\end{equation}
After sending the information $\alpha_{1}$, $\alpha_{2}$ and $\alpha_{3}$ to qubit $4$ by classical communication, the operation $U_{4}(\alpha_{1}, \alpha_{2}, \alpha_{3})$ is performed on qubit $4$, which is defined by
\begin{equation}
U_{4}(\alpha_{1},\alpha_{2},\alpha_{3})=\cos\theta-i\alpha_{1}\alpha _{2}\alpha _{3}\sin\theta\sigma_{4}^{y}.
\label{B.22}	
\end{equation}
The average state after the operation can be given by
\begin{equation}
   \begin{aligned}
   \rho=&\sum_{\alpha_{1},\alpha_{2},\alpha_{3}}U_{4}(\alpha_{1},\alpha_{2},\alpha_{3})P_{3}(\alpha_{3})P_{2}(\alpha_{2})P_{1}((\alpha_{1})) |g\rangle\langle g| \\& P_{1}(\alpha_{1})P_{2}(\alpha_{2})P_{3}(\alpha_{3})U_{4}(\alpha_{1},\alpha_{2},\alpha_{3})^{\dagger}.	
   \end{aligned}
   \label{B.23}	
\end{equation}
The energy extracted from qubits $3$ and $4$ is computed as
\begin{equation}
E_{out}=\frac{1}{\sqrt{4h^{2}+k^{2}}}[3hk\sin2\theta-(2h^{2}+2k^{2})(1-\cos2\theta)].	
\label{B.24}	
\end{equation}
In order to maximize $E_{out}$ with respect to $\theta$, we can obtain the optimal choice for $\theta$, which satisfies
\begin{align}
\cos2\theta & =\frac{2h^{2}+2k^{2}}{\sqrt{(2h^{2}+2k^{2})^{2}+(3kh)^{2}}}, \nonumber\\
\sin2\theta & =\frac{3hk}{\sqrt{(2h^{2}+2k^{2})^{2}+(3kh)^{2}}}.
\label{B.25}
\end{align}
Substituting Eq. $(\ref{B.25})$ into Eq. $(\ref{B.24})$ to obtain the maximum output energy $E_{out}$,
\begin{equation}
E_{out}=\frac{2h^{2}+2k^{2}}{\sqrt{4h^{2}+k^{2}}}[\sqrt{1+\frac{(3kh)^{2}}{(2h^{2}+2k^{2})^{2}}}-1].
\label{B.26}	
\end{equation}
The energy transfer efficiency during this process is calculated as
\begin{equation}	
	\eta =\frac{(2h^{2}+2k^{2})[\sqrt{1+\frac{(3hk)^{2}}{(2h^{2}+2k^{2})^{2}}}-1]}{6h^{2}}.
	\label{B.27}
\end{equation}
\bibliographystyle{apsrev4-1}
\bibliography{ref}

\end{document}